\documentclass[cits,hyper]{pos}

\usepackage{fontenc}
\usepackage{times}
\usepackage{mathptmx}
\usepackage{amssymb}
\usepackage{graphicx}
\usepackage{subfigure}
\newcommand{\tr}{\mbox{Tr}\;}
\newcommand{\M}{\mathbb{D}}
\newcommand{\heplat}[1]{\href{http://arxiv.org/abs/hep-lat/#1}{\texttt{arXiv:hep-lat/#1}}}
\newcommand{\heplatnew}[1]{\href{http://arxiv.org/abs/#1}{\texttt{arXiv:#1 [hep-lat]}}}

\title{First evidence for Casimir scaling in G$_2$ lattice gauge theory%
\addtocounter{footnote}{1}\thanks{This research was supported by the Slovak Grant Agency for Science, Project VEGA No.\ 2/6068/2006. {L\kern-.08cm'}.L.~would like to acknowledge the Organizing Committee of Confinement 8 for support.}}

\ShortTitle{First evidence for Casimir scaling in G$_2$ lattice gauge theory}

\author{\addtocounter{footnote}{-2}\speaker{{L\kern-.08cm'}udov{\'\i}t Lipt\'ak} and {\v{S}}tefan Olejn{\'\i}k\\
        Institute of Physics, Slovak Academy of Sciences, SK--845 11 Bratislava, Slovakia\\
        E-mail: \email{ludovit.liptak@savba.sk}, \email{stefan.olejnik@savba.sk}}

\abstract{Potentials between static quarks and antiquarks from a few lowest representations were evaluated in numerical simulations of 4-dimensional pure G$_2$ lattice gauge theory at various couplings. The obtained potentials are linearly rising at intermediate distances and their string tensions exhibit (approximate) Casimir scaling. This result is in accordance with a model of the vacuum of non-Abelian gauge theories with a domain structure, in which the (color) magnetic flux randomly fluctuates within a domain, but the total flux in each domain is quantized in units of the gauge group center.
}

\FullConference{8th Conference Quark Confinement and the Hadron Spectrum \\
		 September 1-6, 2008\\
		 Mainz, Germany}

\begin{document}

\section{Introduction}\label{sec:intro}

	The behavior of the static quark-antiquark potential in SU($N$) gauge theory in three ranges of interquark distances belongs to basic features of confinement in the case of neglecting dynamical quarks.  For small separations, the interaction is dominated by gluon exchange and the potential is Coulomb-like.  Going to larger distances,  there is an intermediate region with a linearly-rising potential. Numerical simulations in both SU($2$) \cite{Piccioni:2005un} and SU($3$) \cite{Deldar:1999vi,Bali:2000un} lattice gauge theories showed that string tensions  for different representations were approximately proportional to the quadratic Casimir in that region. The phenomenon is called Casimir scaling. 
In the region of color screening, \textit{i.e.}\ at asymptotic distances, the string tensions depend only on the $N$-ality of the representation.
	
	From the point of view of confinement, G$_2$ gauge theory represents an interesting laboratory, and attracted due attention in the last years. The reason is triviality of its center which allows to reassess the role of the group center in the confinement mechanism.  In G$_2$, for example,  a ``quark'' from the fundamental representation can be screened by ``gluons'' (from the adjoint representation), thus the string tension is zero at asymptotic distances. However, the authors of \cite{Greensite:2006sm} showed that the fundamental potential is still linearly rising at intermediate distances.

	To explain linearity, Casimir scaling and asymptotic $N$-ality dependence in a unified way in both SU($N$) and G$_2$ gauge theories, a simple model of the Yang--Mills vacuum was suggested in Ref.\ \cite{Greensite:2006sm}. The model assumes a domain structure of the vacuum where color magnetic fields fluctuate randomly and (almost) independently inside every 2D-slice of the vacuum, \textit{i.e.}\ each 2D-slice is a bunch of small independently fluctuating subregions. On the other hand, the total magnetic flux through each domain corresponds to an element of the center. For example, in G$_2$ all domains will be of the vacuum type because of center triviality. As a prediction, the model leads to Casimir scaling at intermediate distances in G$_2$ gauge theory. In this contribution we present the first evidence for Casimir scaling, more extensive results have meanwhile appeared in Ref.\ \cite{Liptak:2008gx}.

\section{Potential for fundamental representation}\label{sec:fundamental}

	Our lattice implementation of G$_2$ gauge theory is based on the formulation outlined in \cite{Pepe:2005sz}.  One uses a very efficient parameterization of G$_2$ suggested by \cite{Macfarlane:2002hr}. The Wilson action,  given as
\begin{eqnarray}\label{eq:Wilson_action}
S &=& - \frac{\beta}{7} \sum_{x, i > 0} \xi_0 \; \textrm{Re Tr } \left[   P_{i0} (x)  \right]  - \frac{\beta}{7} \sum_{x, i > j > 0} \frac{1}{\xi_0} \textrm{Re Tr } \left[   P_{ij} (x)  \right], 
\end{eqnarray} 
was used. Here $ P_{\nu \tau} (x) $ represents the standard plaquette. We worked on asymmetric lattices of the type $L^3\times(2L)$. The ratio of the lattice spacings in time ($a_t$) and space ($a_s$) directions was fixed by requiring the same total physical length of the lattice in all directions,  \textit{i.e.}\ $\xi=a_s^{\mathrm{phys}}/a_t^{\mathrm{phys}}=2$. The bare anisotropies $\xi_0$, entering the Wilson action (\ref{eq:Wilson_action}) and leading to $\xi=2$, depend on the coupling $\beta$ and were determined following the procedure of Ref.\ \cite{Klassen:1998ua}.

	 The simulations were performed at three values of the coupling $\beta \in \{ 9.5, 9.6, 9.7 \}$. For each coupling we produced almost 1000 configurations. In order to have overlap of our trial quark-antiquark state with the ground state as large as possible, we applied the so-called \textit{stout smearing procedure} of Morningstar and Peardon \cite{Morningstar:2003gk}, properly adjusted for G$_2$. We optimized the smearing separately for each potential.

	The static potential is expressed by means of the expectation values $W(r,t)$ of Wilson loops as
\begin{equation}\label{def_potent}
V(\mathbf{r},a)=-\lim_{\mathbf{t}\to\infty}\frac{1}{\mathbf{t}}\ln W(r,t)
\end{equation}
where $\mathbf{r}=r\cdot a$ is the spatial extent and $\mathbf{t}=t\cdot a$ is the temporal separation of the Wilson loop.  Because of various orientations of the Wilson loop on an asymmetric lattice, one can define three different potentials -- $V_{st}$, $V_{ss}$ and $V_{ts}$.  They are not independent, and each potential $V(\mathbf{r},a) $ consists of two contributions: the true $\mathbf{r}$-dependent static potential due to the interaction, and the self energy contribution (independent of~$\mathbf{r}$). The interaction parts satisfy two simple relations (see \cite{Bali:2000un,Klassen:1998ua}), allowing us to determine the bare-anisotropy parameter of the action.
               
	In accordance with the relation (\ref{def_potent}), the potentials (in lattice units) were determined by fitting logarithms of the measured loops by linear functions in $t$:
\begin{equation}\label{eq:linear_fit}
-\ln W_\lambda(r,t)=C_\lambda+\widehat{V}_\lambda(r)\cdot t,\qquad\lambda\in\{ss, st, ts\},
\end{equation}
in an interval of $t$ values, from $t_\mathrm{min}$ to $t_\mathrm{max}$, where $ \ln W_\lambda(r,t) $ is approximately linear. Using the smearing procedure, logarithms of Wilson loops show a linear behavior down to $t=1$, so   the lower limit was chosen $t_\mathrm{min}=1$. The upper limit of our fits $t_\mathrm{max}$ was not fixed so firmly, and was varied depending on the ground-state overlap. By means of the standard least-square method, we estimated statistical errors of the linear fit (\ref{eq:linear_fit}) in a fixed $(t_\mathrm{min},t_\mathrm{max})$-interval. 
Systematic uncertainties and errors of the procedure were discussed in more detail in Ref.\ \cite{Liptak:2008gx}. 

\begin{figure}[!t]
\centering
    \includegraphics[height=1.8in,width=2.2in]{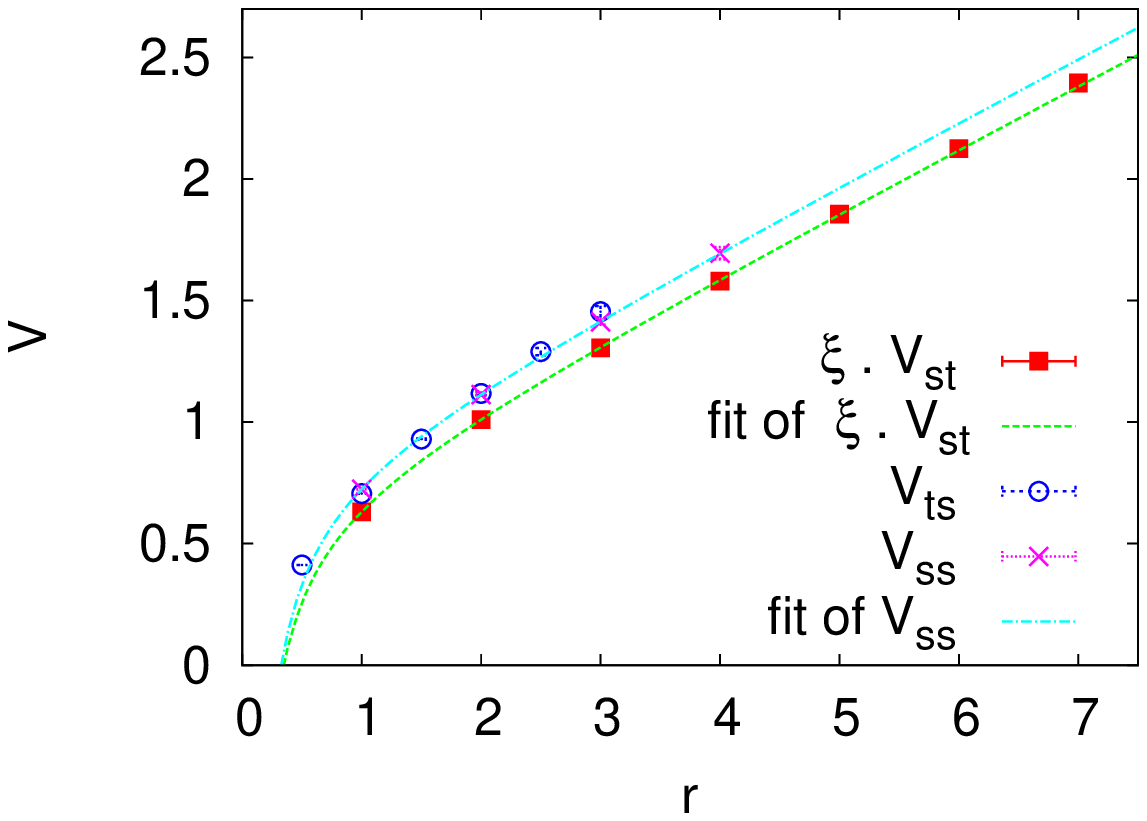}
    \hspace{.3in}
    \includegraphics[height=1.8in,width=2.2in]{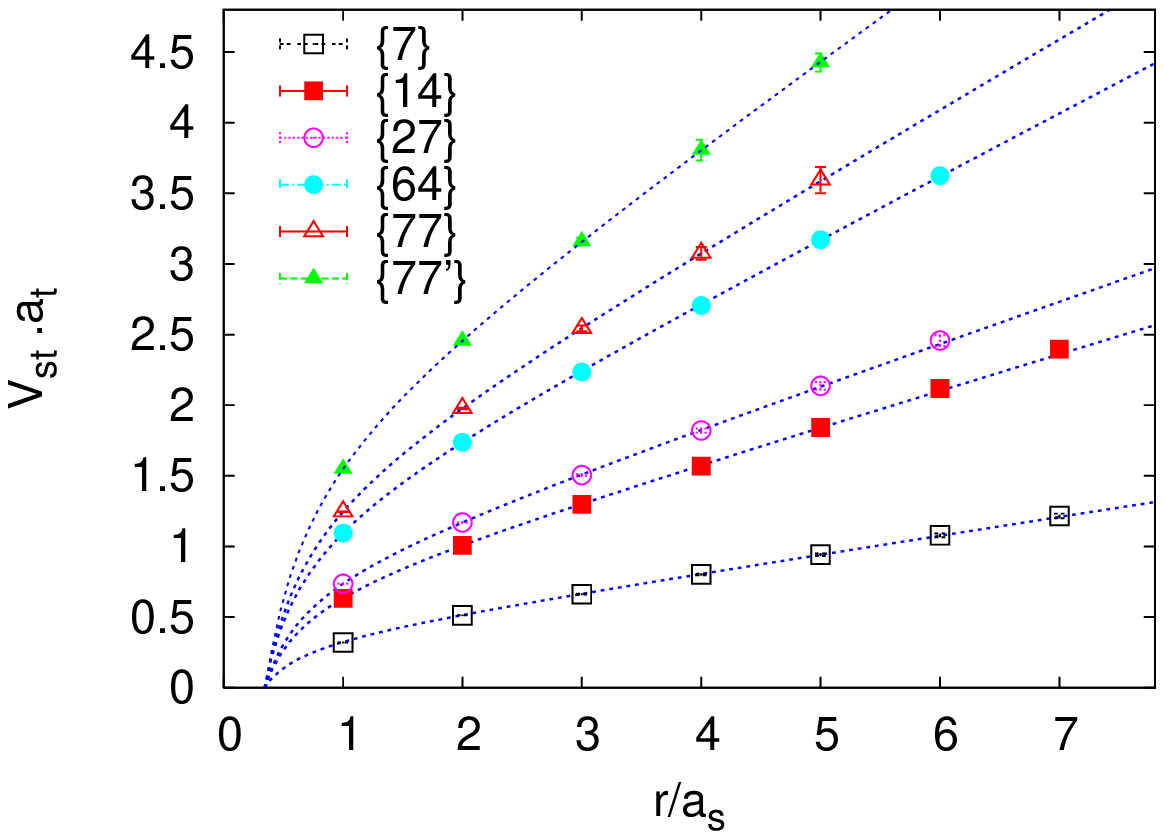}
\caption{(Left:) Potentials $\widehat{V}_{st} $, $\widehat{V}_{ss} $, and $\widehat{V}_{ts} $. Distances $r$ are measured in units of $a_s$. $14^3 \times 28$ lattice, $ \beta = 9.6 $, $\xi = 1.97$. (Right:) Potentials $\widehat{V}_{st, \{ D \}} $ for representations $ \{ D \} $ as functions of the dimensionless $r$. $14^3 \times 28$ lattice and $\beta = 9.6 $ }\label{fig:pot_96plusfit}
\end{figure}

	In the second step, the calculated potentials were fitted by a 3-parameter function, including the Coulomb and linear terms:
\begin{equation}\label{eq:potential_fit}
\widehat{V}_\lambda(r)=\widehat{c}_\lambda-\frac{\alpha_\lambda}{r}+\widehat\sigma_\lambda\; r,\qquad\lambda\in\{ss, st, ts\}.
\end{equation}
For illustration, all three potentials are displayed in Fig.~\ref{fig:pot_96plusfit} (left panel) for $\beta = 9.6$. We can identify easily the range of the linear behavior, and we can also conclude that the potential $ V_{st} $ is most suitable for reliable determination of the string tension. This conclusion is supported by results of the potential-fitting procedure, summarized in Table~\ref{tab:fund_asym}.
	
	Finally, from the numerical values of the string tensions we computed the renormalized aniso\-tropies $\xi= 2.00(7), 1.97(12),$ and $1.96(16)$ for $\beta=9.5, 9.6$, and $9.7$, respectively. These values confirm that the renormalization was done accurately.

\begin{table}[!t]
\begin{center}
\begin{tabular}{c c c c c c c}
\hline \hline
 $ \beta $ & $ \alpha_{st} $   &  $\widehat{\sigma}_{st}$ & $\alpha_{ss}$ &  $\widehat{\sigma}_{ss}$ & $\alpha_{ts}$ &  $\widehat{\sigma}_{ts}$ \\\hline
 9.5 &  0.118(5)   &   0.1997(19) &  0.249(31)  &  0.399(13)   & 0.194(5) &  0.2482(22) \\
 9.6 &  0.124(4)   &   0.1303(18) &  0.264(34)  &  0.257(15)   & 0.231(7) &  0.1784(30) \\
 9.7 &  0.122(5)   &   0.0999(21) &  0.264(36)  &  0.196(15)   & 0.244(8) &  0.1452(33) \\\hline \hline
\end{tabular}
\end{center}
\caption{Parameters of the static potential between fundamental charges. ($14^3 \times 28$ lattice.)}
\label{tab:fund_asym} 
\end{table} 
	
\section{Potentials for higher representations}\label{sec:potentials}

\textbf{\textit{Construction of higher representations:}} Using tensor decompositions of different products of representations, traces of higher-representation matrices can be expressed through traces of the fundamental- and adjoint-representation matrices, \textit{e.g.}\ in the case of the 27-dimensional representation one obtains the relation 
\begin{eqnarray}\label{eq:trace_formulas}
&& {\tr} \M^{\{27\}} = - 1 - {\tr} \M^A - {\tr} \M^F +  ({\tr} \M^F)^2. 
\end{eqnarray}
The adjoint-representation matrix is constructed from the fundamental one:
\begin{equation}\label{eq:fun_to_adj}
\M^A_{ab}(g) = 2\;\tr\left[ \M^F(g)^\dagger \; t_a \; \M^F(g) \; t_b \right],
\end{equation}
where $t_a$ are the generators of the group, based on representation described in \cite{Macfarlane:2002hr}. 

\textbf{\textit{Determination of potentials:}} We focus our attention only on potentials $\widehat{V}_{st}$ which can be calculated most accurately and which exphibit best the linear behavior. On a thermalized, smeared configuration in the fundamental representation, we computed a Wilson loop as a product of link matrices in the fundamental and adjoint representations, and knowing these two traces we calculated also traces of the Wilson loop in higher representations. Finally, we computed their expectation values.
	
	To obtain values of potentials, we use again the linear fit (\ref{eq:linear_fit}). The potentials in higher representations show similar behavior to the case of the fundamental representation -- a linear rise in the same range of distances, but with different slope (see Fig.~\ref{fig:pot_96plusfit}, right panel). 
\begin{table}[!b]
\begin{center}
\begin{tabular}[c]{ c c  c c  c c }  
\hline \hline    
 $\beta$ &  $\widehat\sigma_A/\widehat\sigma_F$ &  $\widehat\sigma_{\{27\}}/\widehat\sigma_F$ & $\widehat\sigma_{\{64\}}/\widehat\sigma_F$ & $\widehat\sigma_{\{77\}}/\widehat\sigma_F$ &  $\widehat\sigma_{\{77 ' \}}/\widehat\sigma_F$   \\\hline
9.5         & 1.88(4) & 2.15(5) &  3.1(1) &    ------       & ------ \\
 9.6         & 1.94(4) & 2.24(6) &  3.35(8)   & 3.8(2)  & 4.6(2) \\
 9.7         & 1.96(6) & 2.28(7) &  3.5(1) & 4.0(2) &  4.9(2) \\ \hline
Prediction& $ 2.0 $     & $ 2.333 $     & $ 3.5 $     & $ 4.0 $     & $ 5.0 $     \\
 \hline \hline
\end{tabular}
\end{center}
\caption{Ratios of string tensions $\widehat{\sigma}_{st}$ for different representations. ($14^3 \times 28$ lattice.)} 
\label{tab:tensions_ratios}
\end{table}

\begin{figure}[!tb]
\label{fig:tensions_ratios}
\begin{center}
\includegraphics[height=1.8in,width=1.9in]{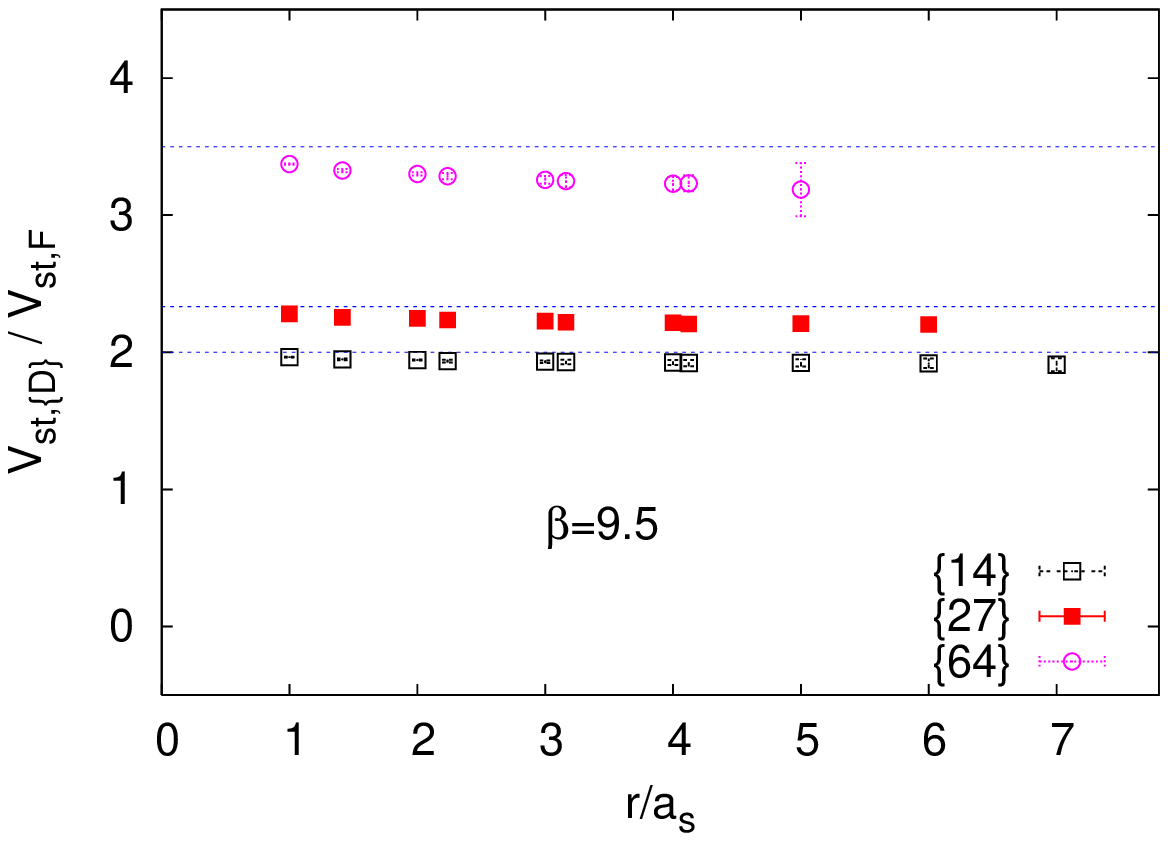} 
\includegraphics[height=1.8in,width=1.9in]{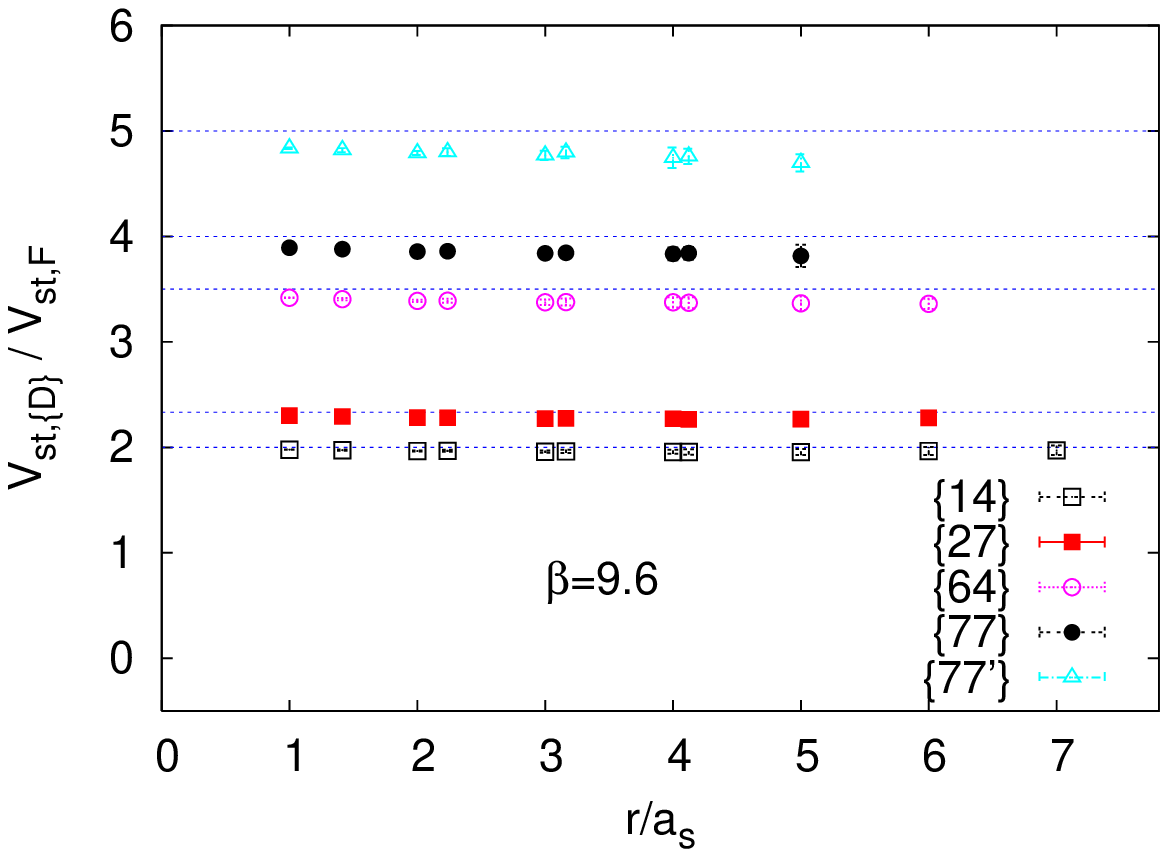} 
\includegraphics[height=1.8in,width=1.9in]{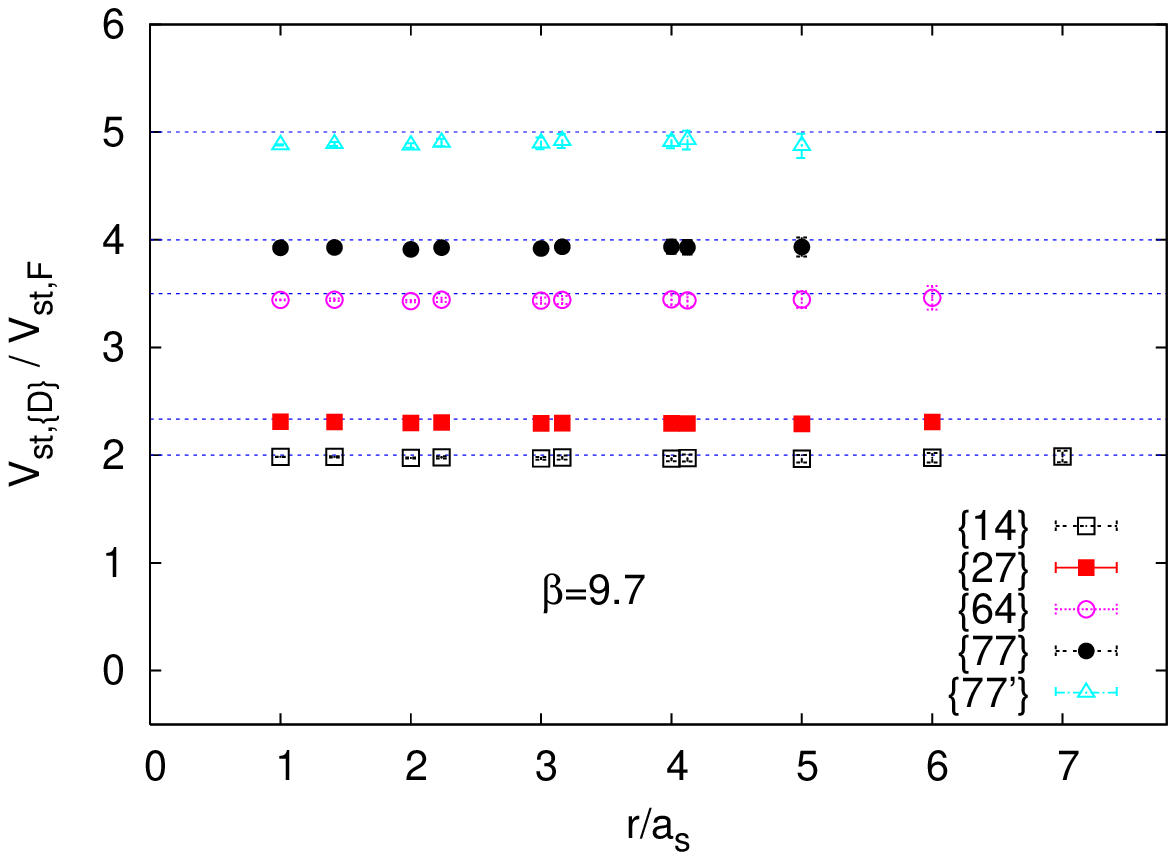} 
\end{center}
\caption{Ratios $\widehat{V}_{st,\{D\}} / \widehat{V}_{st,F}$ for different representations $\{D\}$ are shown as functions of the dimensionless~$r$. Horizontal lines represent predictions of Casimir scaling. ($14^3 \times 28$ lattice.)}\label{fig:ratios}
\end{figure}

	Finally, we fitted the computed potentials again by the 3-parameter fit (\ref{eq:potential_fit}). The resulting ratios of the string tensions are summarized in Table~\ref{tab:tensions_ratios}. The calculated ratios are very close to predictions based on Casimir scaling, and we observe an improvement of the agreement with Casimir scaling by increasing the coupling $\beta$. Fig.~\ref{fig:tensions_ratios}  shows ratios of different-representation potentials to the fundamental one at fixed $r$ values. For each $r$, the ratios barely deviate from the Casimir-scaling prediction.

%

\section{Conclusions}\label{sec:conclusions}

    In the G${}_2$ lattice gauge theory, we calculated string tensions for static potentials between color charges from the six lowest representations of G${}_2$ on asymmetric $L^3 \times (2L)$ lattices.  Ratios of the string tensions in the interval of intermediate distances exhibit Casimir scaling with very good accuracy. The deviations from predictions based on values of quadratic Casimirs were estimated -- including both statistical and systematic errors -- to be at most 10--15\%. The results, together with the evidence for Casimir scaling in SU(2) and SU(3), support the model of magnetically disordered Yang--Mills vacuum with a domain structure.
	
%

\end{document}